\documentclass[aps,pra,twocolumn,floatfix,showpacs]{revtex4-1}
\usepackage{graphicx,amsfonts}
\usepackage{amsmath,amscd,amsfonts,amssymb,color}
\newcommand{\bra}[1]{\left\langle{#1}\right\vert}
\newcommand{\ket}[1]{\left\vert{#1}\right\rangle} 

\begin{document}
\title{A quantum Stirling heat engine operating in finite time}
\author{Debmalya Das$^{1,2,3}$, George Thomas$^{4}$, and Andrew N. Jordan$^{1,2,3}$}

\affiliation{$^1$Department of Physics and Astronomy, University of Rochester, Rochester, New York 14627, USA\\
$^2$Center for Coherence and Quantum Optics, University of Rochester, Rochester, New York 14627, USA\\
$^3$Institute for Quantum Studies, Chapman University, Orange, California 92866, USA\\
$^4$QTF Centre of Excellence, Department of Applied Physics, Aalto University, P.O. Box 15100, FI-00076 Aalto, Finland}

\begin{abstract}
 In a quantum Stirling heat engine, the heat exchanged with two thermal baths is partly utilized for performing work by redistributing the energy levels of the working substance. 
 We analyze the thermodynamics of a quantum Stirling engine operating in finite time. We develop a model in which a time-dependent potential barrier changes the energy-level structure of the working substance. The process takes place under a constant interaction with the thermal bath. We further show that in the limit of slow operation of the cycle and low temperature, the efficiency of such an engine approaches Carnot efficiency. We also show that the maximum output power , for the strokes that affect the energy levels, is obtained at an intermediate operating speed, demonstrating the importance of a finite-time analysis. 
\end{abstract}

\maketitle

\section{Introduction}
In quantum thermodynamics, quantum heat engines~\cite{Scovil1959, Skrzypczyk2014, Qin2017, Li2017, Friedenberger_2017, Kuo2020, Lee2020, Solfanelli2020, Niu2022} and quantum refrigerators~\cite{He2017, Yu2019, Seah2018, Bhandari2021} occupy a central position. As opposed to a classical heat engine, quantum heat engines use quantum matter as their working substance~\cite{Scovil1959, Gemma2004, Quan2007, Deffner2019, Kieu2004, Kieu2006} and operate at length scales where quantum effects are dominant. Quantum heat engines have become test beds for probing the most fundamental questions of quantum thermodynamics such as the definitions of work, heat, efficiency and power in the nanoscale. Quantum analogs of various classical heat engines are being employed to test versions of thermodynamic concepts in the quantum regime~\cite{Thomas2011, Ivanchenko2015, Altintas2015}. They can be employed to investigate the differences between classical and quantum thermodynamic systems and better our understanding of the quantum-classical transition problem of thermodynamic processes~\cite{Quan2006}. The advent of quantum technology has provided a boost to the research in the area of quantum heat engines. Significant strides have been made to better understand how the operations of heat engines and heat cycles in various experimental platforms like trapped ions, quantum dots,  superconducting qubits, etc., depend on quantum effects such as quantum adiabaticity, discreteness of energy levels, quantum statistics, quantum coherence, quantum measurement, and quantum entanglement~\cite{Rosnagel2016, Maslennikov2019, Ono2020, Cimini2020, Benenti2017a, Rosnagel2014, Klatzow2019, Solfanelli2021, Cherubim2019, Jaliel2019}. The investigation of quantum heat engines and quantum refrigerators has great significance in connection to the efficient manipulation and management of heat in microscopic devices and quantum circuits and resetting the state of quantum systems~\cite{Valenzuela2006, Reed2010, Geerlings2013, Bultink2016, Yoshioka2021, Mottonen2021, Guo2022}.

Quantum heat engines can be realized in a variety of ways and can be operated using a variety of cycles~\cite{Deffner2019, Quan2007}. Notable examples of quantum heat engines are the single-atom heat engine with trapped ions~\cite{Rosnagel2016}, the three-level quantum heat engine~\cite{Scovil1959}, the quantum Otto cycle engines~\cite{Thomas2011, Abah2012, Campo2014, Ivanchenko2015, Kosloff2017, Quan2005, Rosnagel2016, Johal2021}, the quantum Carnot heat engines~\cite{Scully2003, Kieu2006, Purwanto2015, Singh2018}, and the measurement-based heat engines~\cite{Elouard2018, Jordan2020, Elouard2020, Anka2021, Bresque2021, Manikandan2022}. Particularly in measurement-based quantum engines, a measurement operation can be used to replace a thermal bath. Certain quantum heat engines, using measurement, can use the measurement result as feedback to update the state of knowledge of the working substance for the next cycle. Interesting examples of such engines are the quantum Maxwell demon engines~\cite{Kieu2006, Kieu2004, Pekola2016, Averin2017} and the quantum Szilard engine~\cite{Kim2011, Mohammady2017, Davies2021, Aydiner2021, Pal2021}.

In a quantum Szilard engine, the work is extracted due to the reorientation of energy levels of a one-dimensional potential box and feedback from a measurement result. The reorientation is achieved by quasistatically inserting and removing a potential barrier in the middle of the potential box while being in equilibrium with a thermal bath. Comparing a stage of the cycle where there is no potential barrier to a stage where the insertion of the barrier is complete, the system transitions to a one-dimensional double-well potential, each well having half the size of the original single well or potential box. As a result, the new energy spectrum is doubly degenerate. A similar idea of extracting work by changing the energy spectrum can be used in a quantum heat engine operating a quantum Stirling cycle~\cite{Thomas2019}. Generally, a quantum Stirling cycle consists of an isothermal compression when in contact with a hot bath, an isochoric thermalization following a change in contact from a hot bath to a cold bath, an isothermal expansion when in contact with a cold bath, and finally an isochoric thermalization following the change in contact from a cold to a hot bath. The version of the quantum Stirling engine exploits the changing degeneracy of the potential box. The isothermal compression and isothermal expansion steps are achieved by the insertion and removal of the potential barrier while in contact with the hot and cold baths respectively, as is done in a quantum Szilard engine. However, unlike the quantum Szilard engine, it does not require the measurement and the feedback steps, and operates with two thermal baths. It has been shown that in the low-temperature and quasistatic limit, the efficiency of the engine approaches the Carnot value ~\cite{Thomas2019}.  Recently, many investigations have been carried out with quantum heat engines based on the degeneracy of the working substance ~\cite{Chattopadhyay2019, Johal2021, Chatterjee2021}. Reference~\cite{Chattopadhyay2019} explores a relativistic quantum Stirling engine, using a potential box as the working substance and bounds the work extraction of the quantum heat engine from the point of view of the uncertainty principle. The bound on work extracted and efficiency is further elaborated in Ref.~\cite{Chattopadhyay2021}. Reference~\cite{Chatterjee2021} realizes a quantum Stirling engine using a harmonic-oscillator potential instead of a potential box.

In  above-mentioned quantum heat engines, the insertion and removal of the potential barrier is necessarily quasistatic. The quasistatic analyses of various thermodynamic cycles are important as they provide limits that serve as a benchmarks to compare the performances of more realistic models. On the other hand, finite-time thermodynamic cycles and open quantum system treatment are essential to find practical estimates of the performance measures of the quantum heat engines. To make the analysis more realistic, in this paper we introduce finite insertion and removal for the barrier with a quantum Stirling engine and separately consider the effect of the thermal bath on the working substance. 

Notably a finite-time quantum Stirling engine has been analyzed in Ref.~\cite{Raja2021}. This model considers a working substance which is a two-level system.  In the quasistatic version of this quantum Stirling cycle, the working substance first undergoes an isothermal compression at temperature $T_h$ during which the energy gap reduces. This is followed by an isochoric thermalization starting with disconnecting the thermal bath at temperature $T_h$ and then connecting to a thermal bath at lower temperature $T_c$. Next an isothermal expansion is carried out at the temperature $T_c$, leading to an increase in the energy gap. The cycle ends with an isochoric thermalization effected by disconnecting the working substance from the thermal bath at temperature $T_c$ and connecting it back to the thermal bath at temperature $T_h$. In particular, the isothermal expansion and compression steps are carried out by means of a driving protocol which can be adjusted to make the processes faster, thus departing from the quasistatic limit and entering the finite-time regime. At the end of the first step a decrease in the energy gap is achieved, but the levels are never degenerate or nearly degenerate.

In this paper we design a quantum heat engine that operates a quantum Stirling cycle in finite time. We take two thermal baths at temperatures $T_h$ and $T_c$ and and put the system in contact with these baths while inserting and removing the potential barrier, respectively. We treat the interaction of the working substance with the heat baths using a Lindbladian master equation, leading to the thermalization. We describe the thin time-dependent potential barrier in the middle of the potential box as a $\delta$-function potential with height changing in time. The effect is equivalent to the drive used in Ref.~\cite{Raja2021}; however, a sufficiently large potential barrier at the end of the insertion step brings pairs of energy levels close enough to be effectively doubly degenerate. 
Another point of difference is that the working substance we use is not a two-level system but a one-dimensional potential well, allowing us to consider a higher number of energy levels. Finally, the dynamics of~\cite{Raja2021} is non-Markovian. On the other hand, as we will see, we consider that in such an interaction with the thermal bath it is possible for the dynamics to be Markovian. Our model is a finite-time version of the quantum Stirling heat engine analyzed in Ref.~\cite{Thomas2019}. To investigate the thermodynamics in the fast regime of insertion of the potential barrier, we consider shorter and shorter times of interaction of the working substance with the thermal bath. We numerically vary a parameter proportional to the length of this time duration of interaction to explore the heat exchanged with the thermal baths and work output in every stroke of the cycle over the entire range of time duration. We show that in the limit of very long time duration of interaction with the thermal baths and for the choice of very low temperatures, the total work output approaches the quasistatic limiting value while the efficiency of the heat engine approaches the corresponding Carnot efficiency. This mirrors the quasistatic behavior of the quantum Stirling engine at low temperatures. In addition, it is shown that the maximum output power for the insertion and removal steps is achieved for a value of efficiency well below the Carnot efficiency value, or in other words, for a speed of driving that does not yield high-efficiency values.

The paper is organized as follows. In Sec.~\ref{sec:model} we present the detailed model of the time-dependent potential barrier and interaction with the thermal baths, necessary for the design of the quantum heat engine. In Sec.~\ref{sec:results} we discuss how the quantum heat engine performs in different regimes of operation rates in terms of work output, energy, and power. We conclude in Sec.~\ref{sec:conc} with a summary and a discussion of potential directions for future work. 

\section{The model}
\label{sec:model}

\begin{figure}
    \begin{center}
    \includegraphics[scale=0.25]{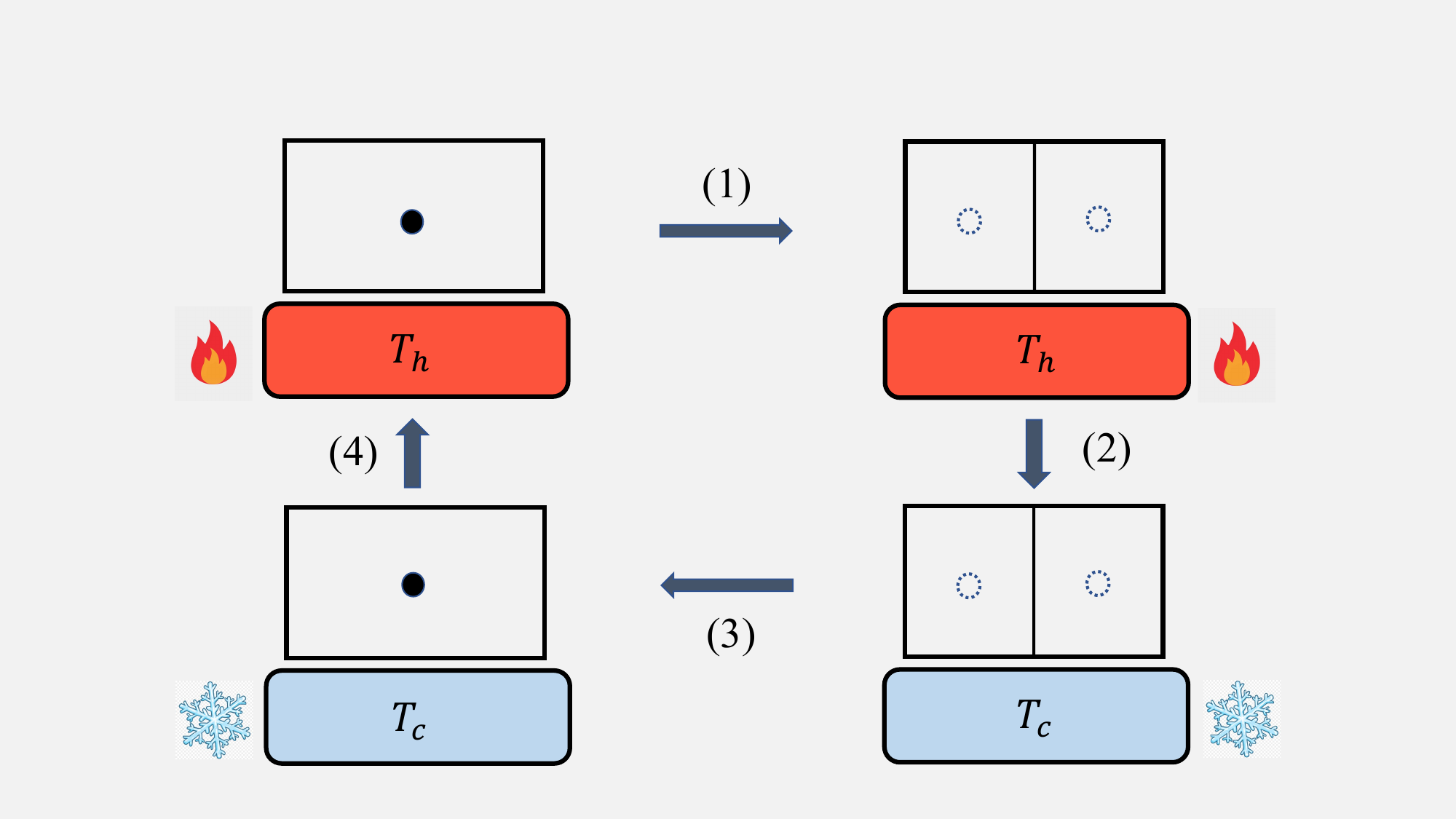}
    \end{center}
    \caption{Four steps of a classical Stirling cycle. In step 1 an isothermal insertion of a barrier is carried out at higher temperature $T_h$. The temperature is changed from $T_h$ to $T_c$, a lower temperature, in step 2. In step 3 the isothermal removal of the barrier is performed at the lower temperature $T_c$. In the last step, step 4, the temperature is changed back to $T_h$. }
    \label{fig:CSC}
\end{figure}

A quantum Stirling cycle is operated between two thermal reservoirs and has four steps (see Fig.~\ref{fig:CSC} for the classical Stirling cycle). Initially, the system is in equilibrium with a thermal bath at temperature $T_h$. As a first step of the cycle, a potential barrier is quasistatically inserted in the middle of the potential box, keeping the temperature fixed at $T_h$. After the insertion of the potential barrier is complete, the temperature is changed by removing the high-temperature thermal bath and bringing the system to equilibrium with a thermal bath at a lower temperature $T_c$. In the next step, the potential barrier is quasistatically removed, keeping the temperature fixed at $T_c$. Finally, the thermal bath at temperature $T_c$ is removed and the system is again brought to equilibrium with the higher-temperature thermal bath $T_h$. At very low temperatures and/or with boxes of sufficiently small lengths, the efficiency of such a quantum heat engine approaches the Carnot efficiency~\cite{Thomas2019}.

Now the quasistatic operation of the quantum Stirling cycle is a highly idealized situation and should take infinite time to complete. Our main motivation is to design a quantum heat engine that works exclusively on quantum features but at the same time is close to practical realization. To perform the process in practice, we thus require an understanding of the processes of insertion and removal of the potential barrier in finite time. An analysis of how the energy levels shift and change with a $\delta$-function potential barrier of growing height has been done in Ref.~\cite{Sordal2019}. However, this result is not translated to understanding the thermodynamics of the complete cycle and how the physics changes and what interesting effects appear over different speeds of operation in terms of insertion and removal of the potential barrier. In addition, it is very important to note that these processes take place under a constant interaction with thermal baths and the working substance is constantly subjected to thermalization. Therefore any analysis of these processes must necessarily include the added interaction of the thermal baths treated with the help of techniques of open quantum system dynamics.

To describe a quantum heat engine operating a quantum Stirling cycle in finite time, we need to model the insertion and removal of the barrier in a way that is consistent with quantum mechanics. In addition to this, we must deal with the interaction of the heat engine with the thermal baths accordingly. A Stirling cycle begins with a particle (or many particles) in a chamber in equilibrium with a heat bath at a temperature $T_h$. The chamber is impervious to any particle but is able to exchange heat with the thermal bath. To realize the quantum version of the cycle, the chamber is replaced by a one-dimensional potential box and a potential barrier in the form of a $\delta$-potential of increasing or decreasing height, with time, is placed in the middle to simulate the insertion and removal of the partition, respectively.

\begin{figure*}
    \centering
    \includegraphics[scale=0.5]{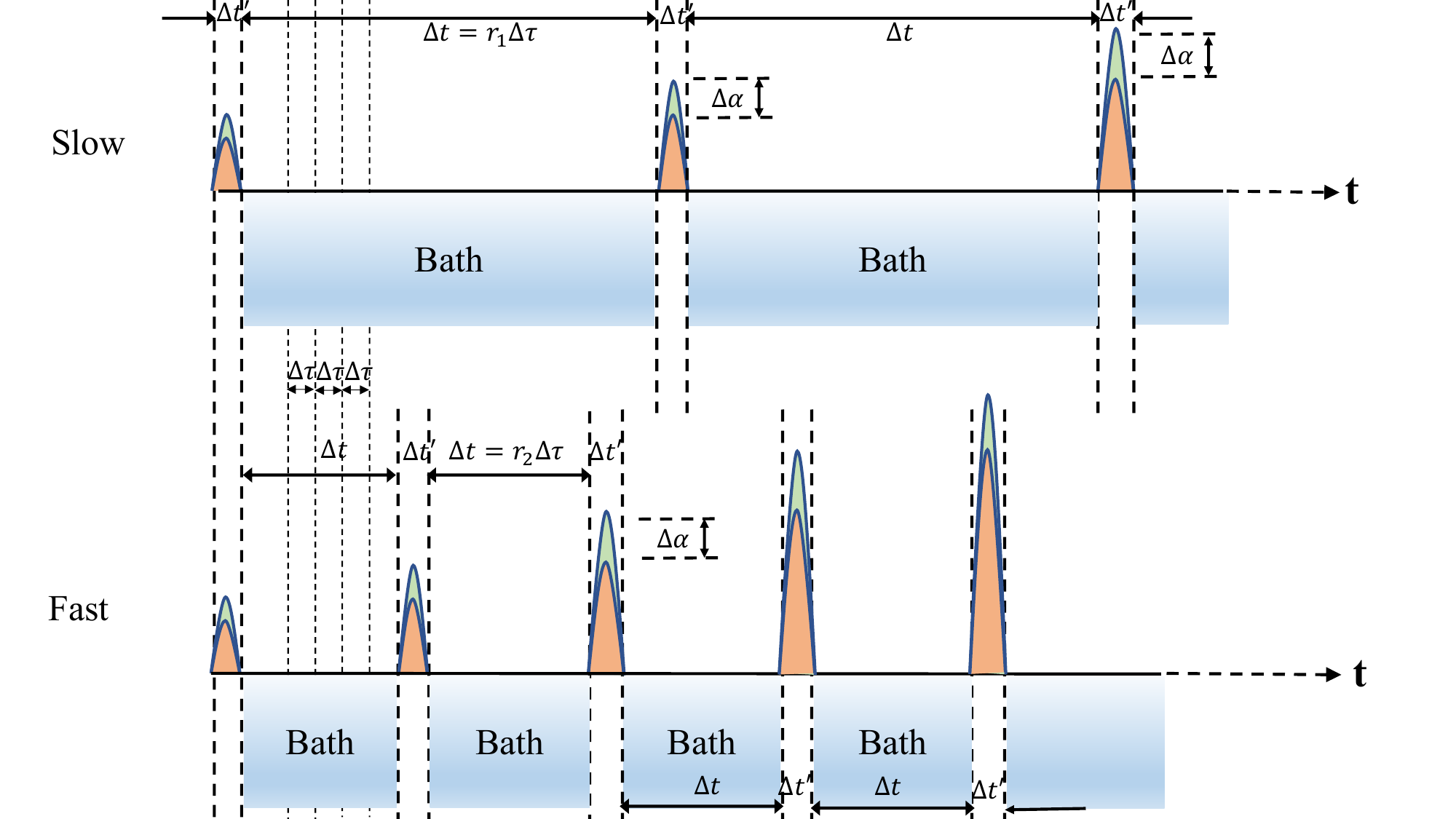}
    \caption{The slow vs fast process in a quantum Stirling engine: Every time-step begins with an adiabatic increase $\Delta \alpha$ of the height of the potential barrier in a duration $\Delta t^\prime$. Eventually we consider that $\Delta t^\prime \rightarrow 0$, so that this adiabatic increase is instantaneous. After this the working substance is in contact with a thermal bath for a duration $\Delta t  = r\Delta \tau$. For a slow process $(a)$, $r$ is considerably larger than for a fast process $(b)$. The quasistatic limit is achieved for $r\rightarrow \infty$. }
    \label{fig:my_label}
\end{figure*}

At any time $t$, the potential profile of the system for a potential box of length $2a$ with a barrier in the middle is given by
\begin{eqnarray}
V(x,t)&=&\alpha(t)\delta(x)\;\; -a<x<a\\
      &=&\infty,\;\;\;\;\;\;\;\; x<-a,\; x>a,\nonumber
\end{eqnarray}
with $\alpha(0)=0$ in the case of insertion. We now explain how we describe the speed of operation of the cycle. By speed of operation we refer to how rapidly the potential barrier is inserted or removed.
We can choose a constant increase $\Delta \alpha$ in the height of the potential barrier corresponding to each time step $\Delta t$.
In this approach, we consider that the insertion or removal of the barrier consists of adiabatic and thermalization steps. The time taken for each adiabatic step is $\Delta t^\prime$, during which $\alpha\rightarrow\alpha+\Delta \alpha$. After each adiabatic step, the system is in contact with the bath and thermalization takes place for $\Delta t=r\Delta \tau$ time. Thus each thermalization step following an adiabatic change of the potential barrier height is made up of $r$ elementary thermalizations of equal duration $\Delta\tau$. We consider the limit $\Delta t'\rightarrow 0$, i.e., a sudden process. Therefore the total time taken for the composite step is $r \Delta \tau$. The quasistatic limit is achieved when $r\rightarrow \infty$ with finite $\Delta t=r \Delta \tau$. This scenario is equivalent to the case where the system is in contact with thermal bath throughout and $r$ discrete steps, during which the height of the barrier changes as $\alpha\rightarrow\alpha+\Delta \alpha$ and the interval between the discrete steps is $\Delta \tau$. 
A large $\Delta t$ would then indicate a slowly driven process leading to a longer interaction with the thermal bath and a longer time to thermalize, with $\alpha$ remaining constant through the duration of the thermalization. The reasoning can be applied accordingly for a fast process. Figure~\ref{fig:my_label} helps to better understand these representations and the relationships between the time durations and the difference between a fast and a slow process.

The adiabatic insertion of a potential barrier in a potential box and thereby splitting the wave function into two parts has been treated extensively in Ref.~\cite{Sordal2019}. The exercise is essentially solving for the energy eigenfunctions and eigenvalues of a potential box with a time-dependent potential barrier. Therefore, we use these results to treat the insertion as well as the removal of the barrier. We consider the potential box to be extending from $-a$ to $+a$, in the $x$ direction, and locate the time-dependent potential barrier at $x=0$. In this approach, the time-dependent Hamiltonian of a particle of mass $m$ corresponding to the insertion of the potential barrier is given by
\begin{equation}
\label{eq:Hamiltonian}
    H(x,t)=-\frac{\hbar^2}{2m}\frac{\partial^2}{\partial x^2}+\alpha(t)\delta(x).
\end{equation}
If $\ket{\psi_n(x,t)}$ is the instantaneous eigenfunction of the Hamiltonian $H(t)$, corresponding to the eigenvalue $E_n(t)$,
\begin{equation}
\label{eq:eveq}
    H(x,t)\ket{\psi_n(x,t)}=E_n(t)\ket{\psi_n(x,t)}.
\end{equation}
In the case of a potential box in one dimension of length $2a$ with the potential barrier of increasing height being introduced in the middle, using Ref.~\cite{Sordal2019}, the wavefunctions corresponding to the instantaneous eigenvectors are given by
\begin{eqnarray}
\label{eq:ef}
\psi_n(x,t) &=& A_n(t) \sin k_n(t)[x+a],\;\;\;-a<x<0\nonumber\\
            &=& B_n(t) \sin k_n(t)[x-a],\;\;\;\;\;\;\;0<x<a,
\end{eqnarray}
where $k_n(t)=\frac{\sqrt{2mE_n(t)}}{\hbar}$ and 
\begin{equation}
\label{eq:nc}
    A_n(t)=\left(a-\frac{\sin{2k_n(t)a}}{2k_n(t)}\right)^{-1}.
\end{equation}
For odd $n$, we have $B_n(t)=-A_n(t)$ and for even $n$ we have $B_n(t)=A_n(t)$. Note here that
\begin{eqnarray}
    \int_{-a}^0 |\psi_n (x,t)|^2 dx= \frac{1}{2}\nonumber\\
    \int_{0}^a |\psi_n (x,t)|^2 dx= \frac{1}{2}
\end{eqnarray}

We now bring the thermal bath into the picture. Consider the system to be in equilibrium with a thermal bath at temperature $T$. A useful description of such a system is the thermal state $\rho$, in which the population of each energy level of the potential box depends on the temperature of the system, in this case the bath-temperature. Accordingly, at the beginning of the cycle,
\begin{equation}
\label{eq:thst}
    \rho(0)= \frac{e^{-\beta H(0)}}{{\rm Tr}[e^{-\beta H(0)}]}=\frac{1}{Z_0}\sum_n e^{-\beta E_n(0)}\ket{\psi_n(0)}\bra{\psi_n(0)},
\end{equation}
where $\beta=\frac{1}{k_B T}$ and $Z_0=\sum_n e^{-\beta E_n(0)}$ is the canonical partition function of the system.

As the potential barrier is changed, the energy spectrum of the system changes. Heat is exchanged with the bath and work is performed. Let us understand the isothermal insertion of the potential barrier. Starting from an initial state where there is no barrier, as the barrier is introduced, the odd numbered energy levels begin shifting towards the next higher even numbered energy levels that remain stationary. This can be very easily seen from~Eq. (\ref{eq:ef}). The even numbered wavefunctions have nodes at $x=0$ throughout the entire process and hence do not undergo any change in their shapes. On the other hand, the odd numbered wavefunctions have anti-nodes at $x=0$ and consequently, with the increase in the height of a potential barrier, dips appear in the wavefunction and then grow at that point, culminating in the points being transformed into nodes when the barrier reaches infinity. The previously odd energy levels now coincide with their next higher even energy levels and the new energy levels are doubly degenerate. At this point it is important to clarify what we mean by the height of the potential barrier to be infinity. For very large values of $t$, the potential barrier is very high and the time required for tunneling of a particle from one side of the well to the other is significantly higher than the time of thermodynamic processes. So for all practical purposes we consider that after such a long time, the height of the barrier is infinity and the insertion is complete. In other words, the single infinite potential well has now been converted to a double well. 

As the insertion of the barrier is performed with the system in contact with a thermal bath, the system is now an open quantum system and cannot be described by closed system dynamical equations. The Schr\"odinger von Neumann equation is no longer sufficient for such situations. We use Lindblad's master equation to study the dynamics of such systems,
\begin{widetext}
\begin{equation}
    \Dot{\rho}=-\frac{i}{\hbar}[H,\rho]+\sum_k \gamma_k(N_{\omega_k}+1) \left[L_k \rho L_k^\dag - \frac{1}{2}(L_k^\dag L_k \rho + \rho L_k^\dag L_k)\right]+
    \sum_k \gamma_k N_{\omega_k} \left[L_k^\dag \rho L_k - \frac{1}{2}(L_k L_k^\dag \rho + \rho L_k L_k^\dag)\right],
\label{eq:master_eq}    
\end{equation}
\end{widetext}
where the set of operators $\{L_k\}$ are called jump operators or Lindblad operators and are given by $L_k=\ket{\psi_{k-1}(t)}\bra{\psi_{k}(t)}$. Essentially, they cause transitions from the $k$th to the $(k-1)$th energy level in the potential well. Note that the Lindblad operators are defined with respect to instantaneous eigenbasis of the Hamiltonian at every step lasting time $\Delta t$. Here $\gamma_k$ is the decay rate corresponding to the transition $L_k$. In this paper we work with an Ohmic bath so that $\gamma_k \propto \Delta \omega_k$ where $\Delta\omega_k=\frac{E_k-E_{k-1}}{\hbar}$~\cite{Thomas2019a, Breuer2002, Breuer2016}. Here $N_{\omega_k}=\frac{1}{e^{\beta \hbar \Delta\omega_k}-1}$ represents the Bose-Einstein distribution function for the angular frequency transition $\Delta\omega_k$. The first term containing the commutation of the system Hamiltonian and the density operator gives the unitary evolution. The second and third terms are the results of interactions with the thermal bath accompanied by transitions from an energy level to the next lower level and vice versa, respectively. 

Note that Eq. (\ref{eq:master_eq}) is Markovian. This can be ensured in the following way. The decay rate should be such that $\gamma_k \propto g^2$, where $g$ is the coupling between the system and the bath. In the regime $g \ll \Delta \omega_k$, the coupling is weak and the process is Markovian~\cite{Breuer2002, Breuer2016}. The point to be noted is that at certain times, say at the end of the first step or at the beginning of the third step of the quantum Stirling cycle, $\Delta \omega_k \rightarrow 0$. This corresponds to the limit when pairs of energy levels overlap or nearly overlap and result in near degeneracy. In this limit of $\Delta \omega_k \rightarrow 0$, the process is no longer Markovian. However, the contribution of this limit towards work output is negligible and hence the process remains practically Markovian for the entire duration of the insertion and the removal steps.

We now briefly recapitulate how we intend to enforce the speed of a process. As discussed earlier, if for a constant value of barrier height $\alpha$, the time duration $\Delta t$ for which the system interacts with the thermal bath is longer, then the process is slow. However, if for a constant $\alpha$, $\Delta t$ is smaller, then the process is fast. As defined earlier, the time duration is $\Delta t=r\Delta \tau$. Here $\Delta \tau$ is an elementary constant time duration and $r$ is a parameter that determines the speed of operation of the cycle. Thus a larger $r$ means a larger $\Delta t$ and hence a slower process and so on. 

Suppose at any given time $t$ the state of the system is
\begin{equation}
    \rho(t)=\sum_n \sum_m p_{nm}\ket{\psi_n(t)}\bra{\psi_m(t)},
\end{equation}
with $p_{mn}\in \mathbb{C}$. The instantaneous height of the potential barrier is $\alpha$ and the  Hamiltonian of the system at this stage is given by
\begin{equation}
    H(t)=\sum_n E_n(t)\ket{\psi_n(t)}\bra{\psi_n(t)}.
\end{equation}
At the beginning of the next time step, there is a sudden change $\alpha \rightarrow \alpha+\Delta \alpha$ and then $\alpha$ remains constant for the entire duration $\Delta t$. With the change in $\alpha$, the energy eigenvalues and the corresponding eigenfunctions shift. Consequently, there is also a change in the Lindblad operator $L_k$.  The new Hamiltonian of the system is given by
\begin{equation}
    H(t+\Delta t)=\sum_n E_n(t+\Delta t)\ket{\psi_n(t+\Delta t)}\bra{\psi_n(t+\Delta t)}.
\end{equation}
At the end of the first elementary time $\Delta \tau$, with $\Delta \tau\rightarrow 0$, the density operator updates according to
\begin{equation}
    \rho(t+\Delta \tau)=\rho(t)+\dot{\rho}(t)\Delta \tau,
\end{equation}
and for the entire time-step $\Delta t$ this gives
\begin{equation}
    \rho(t+\Delta t)=\rho(t)+\sum_{m=0}^{r-1}\dot{\rho}(t+m\Delta \tau)\Delta \tau.
\end{equation}
In any time-step, the heat exchanged with the thermal bath is given by
\begin{equation}
    \Delta Q(t,t+\Delta t)=Tr[H(t)(\rho(t+\Delta t)-\rho(t))]
    \label{eq:heat_in-rem}
\end{equation}
while the work done is 

\begin{equation}
    \Delta W(t,t+\Delta t^\prime)=Tr[(H(t+\Delta t^\prime)-H(t))\rho(t)].
    \label{eq:work_in-rem}
\end{equation}

To obtain the heat exchanged or work done for a complete step of the cycle such as insertion or removal of the potential barrier, one simply has to add the contributions of each time-step as given by~Eqs.(\ref{eq:heat_in-rem})
 and~(\ref{eq:work_in-rem}). Therefore, we have
 \begin{eqnarray}
     Q_{ins}&=&\sum_{i=1}^{n_{\Delta t}} \Delta Q(t_i,t_i+\Delta t)\nonumber\\
     W_{ins}&=&\sum_{i=1}^{n_{\Delta t}} \Delta W(t_i,t_i+\Delta t),
 \end{eqnarray}
 where the $n_{\Delta t}$ represents the number of time-steps required to introduce the potential barrier for a given increase $\Delta \alpha$, or vice-versa for removal of the barrier.
 
 As discussed earlier, to demonstrate our point using the example of the insertion of the potential barrier, the process is complete when the barrier height is infinity. This means that the barrier height must be large enough so that the time scales of all thermodynamic operations, such as changing the barrier height, are much shorter than the time required for any tunneling of a particle. Another way to determine whether the insertion of the potential barrier is complete is to check whether an odd energy level coincides with or is sufficiently close to the next higher even energy level. Similar reasoning can be used for the step involving the removal of the potential barrier, only in reverse. 
 
 We must also take into account the heat exchanged in the other two steps of the cycle, in which the working substance is disconnected from one thermal bath at temperature $T_h(T_c)$ and brought into equilibrium with the other bath at temperature $T_c(T_h)$. Let us take one of the cases. When the bath at temperature $T_h$ is removed and the system is allowed to go to equilibrium with the bath at temperature $T_c$, the initial state is a non-equilibrium state $\rho_h(t_{ins})$ with a configuration of eigenstates $\{\ket{\psi_n(t_{ins})}\}$ achieved at the end of barrier insertion at time $t_{ins}$ when they correspond to energy eigenvalues $\{E_n(t_{ins})\}$. 
 When the system goes into equilibrium with the thermal bath at the lower temperature $T_c$, the new state is a thermal state given by
 \begin{equation}
     \rho_c(t_{hc})=\frac{1}{Z(T_c)}\sum_n e^{-\beta(T_c)E_n(t_{ins})}\ket{\psi_n(t_{ins})}\bra{\psi_n(t_{ins})},
 \end{equation}
 where $Z(T_{h,c})=\sum_n e^{-\beta(T_{h,c})E_n}$, $\beta(T_{h,c})=\frac{1}{k_B T_{h,c}}$ and we have assumed that the system goes into equilibrium with the bath at temperature $T_c$ at a time $t_{hc}$. Note during this stroke of the cycle, the energy eigenvalues and eigenstates do not change. The heat lost to the bath in this step is then given by
 \begin{equation}
     Q_{hc}=Tr[H(t_{ins})(\rho_c(t_{hc})-\rho_h(t_{ins}))].
 \end{equation}
 We can use a similar logic to calculate the heat $Q_{ch}$ absorbed by the system at for the next interchange of the thermal baths after the removal of the potential barrier.
 
 The heat exchanged and the work done by the quantum Stirling engine is now straightforward.
 \begin{eqnarray}
     Q&=&Q_{ins}+Q_{hc}+Q_{rem}+Q_{ch},\nonumber\\
     W&=&W_{ins}+W_{rem}.
 \end{eqnarray}
 The efficiency is given by
 \begin{equation}
     \eta=1+\frac{Q_{hc}+Q_{rem}}{Q_{ch}+Q_{ins}}.
 \end{equation}
 The expression for power for the steps involving insertion and removal of the potential barrier is 
 \begin{equation}
     P=\frac{W}{2r n_{\Delta t}\Delta \tau}.
 \label{eq:power_f}    
 \end{equation}
 
As a reminder, first we consider that the system is in equilibrium at the beginning of the strokes involving insertion and removal of the potential barrier. Second, the expression for power in Eq.~\ref{eq:power_f} does not take into account the time taken by the non equilibrium states at the end of insertion or removal of the potential barrier to go into equilibrium following changes in contact with thermal baths, corresponding to the isochoric strokes. In the subsequent section, we evaluate the performance of the quantum Stirling engine in different regimes of operation, in terms of the speed of operation, using numerical simulations.
 
\section{Numerical simulations and performance of the finite time Stirling engine}
\label{sec:results}
In this section we investigate the extractable work, efficiency and power of the quantum heat engine we have modelled, at low temperatures. The numerical simulations are performed for thermal baths at temperatures $T_h=0.1K$ and $T_c=0.05K$ and considering the first four levels of the potential well so that $n_{max}=4$.
The elementary time step $\Delta \tau$ is chosen to be very small, $\frac{2\pi\hbar}{10000(E_4-E_3)}$. The half-length of the potential box $a=20$ nm and $m=m_e$, the mass of electron. The decay rate for the $k$th transition is chosen as $\gamma_k= \frac{\Delta \omega_k}{50}$ to ensure a Markovian evolution as discussed earlier. For the first set of study, we choose $\Delta \alpha=\frac{E_1}{\sigma}$, with $\sigma=50$, and recall that, $\Delta \alpha$ is the change in the height of the potential barrier in the vanishing time $\Delta t^\prime$. In the next set of study, data are generated using $\sigma=100$ so that during insertion, in each vanishing time $\Delta t^\prime$, the increase in the height of the potential barrier is much smaller.

\begin{figure}
\begin{center}
\includegraphics[scale=0.9]{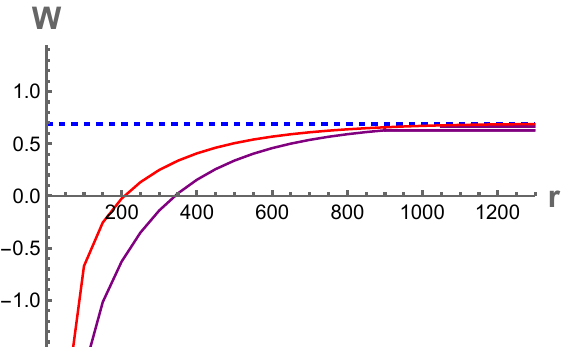}
\begin{picture}(0,0)
\put(-97,150){\boldmath{\Large{$/k_B T_C$}}}
\end{picture}
\end{center}
\caption{Plot of work output $W$, in units of $k_B T_c$, vs cycle speed $r$ for $T_h=0.1$ K and $T_c=0.05$ K. The speed of the operation decreases as $r$ increases. The plot can be divided into two regimes, corresponding to the different signs of $W$. For a very fast driving regime, $W$ is negative and the system does not function as a heat engine. In the regime where $W$ is positive, useful work is extracted and the system functions as a heat engine. For very slow operation $W$ approaches the value $k_B (T_h-T_c){\rm ln} 2$ (blue dotted line) which is the value of work output for a quantum Stirling engine driven quasistatically. The purple and red plots represent different changes $\Delta \alpha=\frac{E_1}{50}$ and $\Delta \alpha=\frac{E_1}{100}$, respectively, of the potential barrier in the vanishing time for $\Delta t^\prime$.}
\label{fig:Work-speed}
\end{figure}
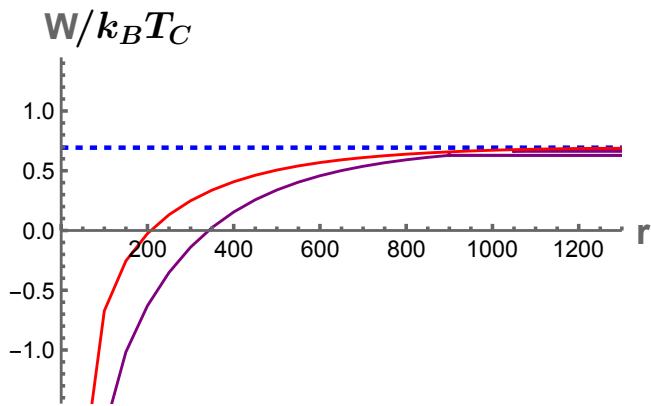

It is interesting to study work output of a finite time quantum Stirling engine in different regimes of operation. As discussed before, we control the speed of operation by varying the time of interaction of the working substance with the thermal bath for a constant value of the height of the potential barrier. Therefore the higher this interaction time, the slower the process. We plot the work output with the parameter $r$ which is proportional to the duration of the time-step for the two values of $\Delta \alpha$ (see Fig.~\ref{fig:Work-speed}).

Clearly the behavior can be divided into two separate regimes. For very small values of $r$, the speed of driving is very fast and the work output is negative. This indicates that the system heats up as a result of internal friction. In this region the system does not function as a heat engine. In the regime where work extracted is positive, the system functions as a heat engine and for large values of $r$, the work output asymptotically tends to a limiting value $k_B(T_h-T_c)\ln2$ which is the low temperature case for quasistatic speed of operation~\cite{Thomas2019}. The plot for the larger value of change in the height of the potential barrier in vanishing time $\Delta t^\prime$ i.e. $\Delta \alpha$, is shifted to the right with respect to the smaller value plot. The larger $\Delta \alpha$ requires a larger value of $r$ or a longer time to interact with the heat bath to reach the regime where useful work can be extracted.

\begin{figure}
\begin{center}
\includegraphics[scale=0.9]{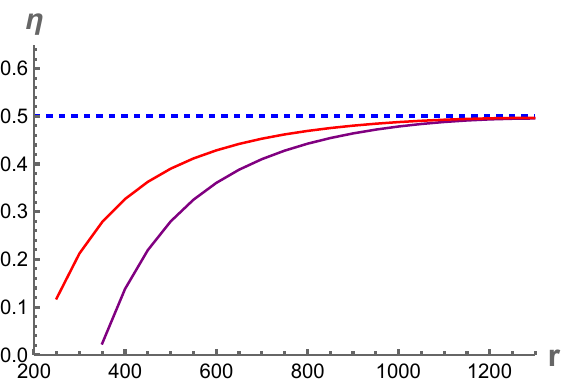}
\end{center}
\caption{Plot of efficiency $\eta$ vs cycle speed $r$ for $T_h=0.1K$ and $T_c=0.05K$. Here $\eta$ increases with $r$, which means the quantum Stirling engine is more efficient when operated slowly. In very slow operating regime, $\eta$ asymptotically tends to the Carnot efficiency value of 0.5 (blue dotted line). The purple and red plots correspond to $\Delta \alpha=\frac{E_1}{50}$ and $\Delta \alpha=\frac{E_1}{100}$, respectively, which are different changes of the potential barrier in vanishing time $\Delta t^\prime$.}
\label{fig:eff-speed}
\end{figure}
 
We witness this behavior once again for efficiency of the quantum Stirling engine as a function of $r$  as shown in Fig.~\ref{fig:eff-speed}. In the limit of very slow driving or large values of cycle speed $r$, the efficiency of the quantum heat engine approaches the Carnot efficiency at the corresponding temperatures. This aspect of the quasistatically driven quantum Stirling engine at low temperatures was shown in Ref.~\cite{Thomas2019}. The dependence on $\Delta \alpha$ matches our intuition discussed regarding the $W$ vs $r$ plot.

\begin{figure}
\begin{center}
\includegraphics[scale=0.9]{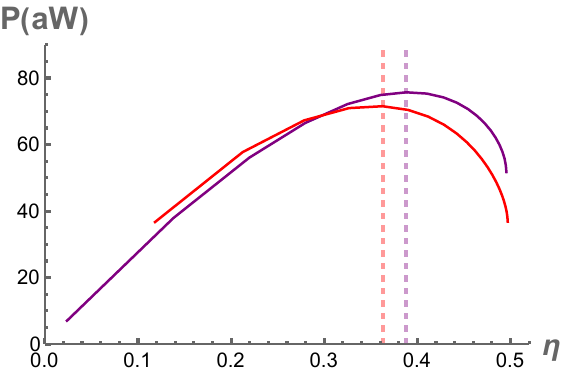}
\end{center}
\caption{Plot of power $P$ in attowatts vs efficiency $\eta$. The maximum $P$ is reached for an intermediate value of $\eta$, much lower than the limiting value of Carnot efficiency at quasistatic rate. For $\Delta \alpha=\frac{E_1}{50}$ (purple solid line), $P_{max}=75.71$ aW corresponds to $\eta=0.39$ (purple dashed line). The maximum power $P_{max}=71.53$ aW is lower for $\Delta \alpha=\frac{E_1}{100}$ (red solid line) and this also corresponds to a lower value of efficiency $\eta=0.36$ (red dashed line) of the heat engine.}
\label{fig:p-eff}
\end{figure}

\begin{figure}
\begin{center}
\includegraphics[scale=0.9]{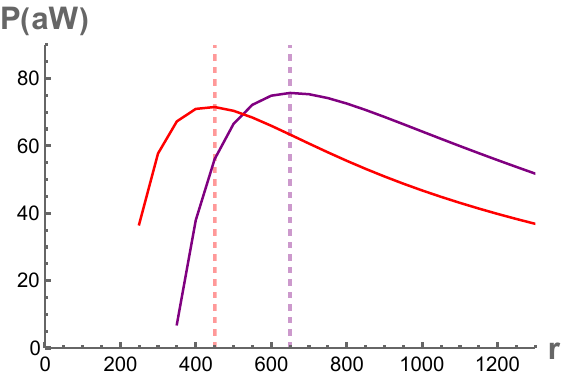}
\end{center}
\caption{Plot of power $P$ in attowatts vs $r$. The maximum $P$ is achieved for $r=650$ (purple dashed line) in the case of $\Delta \alpha=\frac{E_1}{50}$ (purple solid line) and for $r=450$ (red dashed line) in the case of $\Delta \alpha=\frac{E_1}{100}$ (red solid line). The corresponding maximum values of power are $75.71$ aW and $71.53$ aW, respectively. This indicates that the quantum Stirling engine gives the highest power at an operating speed which is neither too fast nor too slow, for a given choice of $\Delta \alpha$.}
\label{fig:p-speed}
\end{figure}

Another property of the quantum Stirling engine driven in finite time is seen from the plot of power, for the insertion and removal steps, as a function of efficiency in Fig.~\ref{fig:p-eff}. We point out once more that at the beginning of these steps, the working substance is considered to be in equilibrium with the respective thermal bath. The time required for the non equilibrium to equilibrium state transition is not considered in the expression for power. Interestingly, the maximum power of the engine is not achieved at maximum efficiency. We have already seen that maximum efficiency is reached in the limit of quasistatic speed of operation or very large $r$ (see Fig.~\ref{fig:p-speed}). However, power reaches its maximum at a lower value of cycle speed $r$, i.e., at an intermediate speed of operation. This important aspect is a different result, not in previous treatments of the quantum Stirling engine done previously. The maximum power achieved for a change in potential barrier $\Delta \alpha=\frac{E_1}{50}$ in time $\Delta t^\prime$ is greater than that for $\Delta \alpha=\frac{E_1}{100}$. Also the value of efficiency corresponding to the maximum power for the former is greater than in the other case.

Clearly, to understand the operation of the quantum Stirling heat engine, we need to appreciate the interplay of the parameters $\Delta \alpha$ and $r$. The larger value of $r$ corresponds to a longer time available to thermalize and hence a slower process. The maximum output powers in the two cases show that most efficient operation does not yield maximum power. 

\begin{figure}
\begin{center}
\includegraphics[scale=0.85]{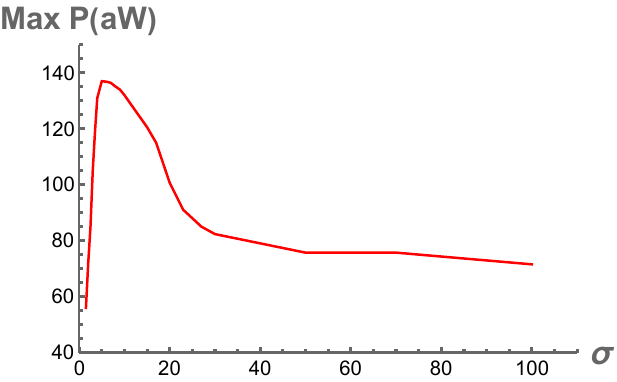}
\end{center}
\caption{Plot of maximum power output vs $\sigma$. A lower value of $\sigma$ indicates a greater change $\Delta \alpha$ in the barrier height in time $\Delta t^\prime$. As $\sigma$ is lowered from 100, we first see an increase in the maximum output power and then a sharp decline.  }
\label{fig:max-p-speed}
\end{figure}

To better understand the role of $\Delta \alpha$, the change in the potential barrier in the vanishing time-step $\Delta t^\prime$, we analyze the plot of maximum power with the ratio $\sigma= \frac{E_1}{\Delta \alpha}$ as shown in Fig.~\ref{fig:max-p-speed}.  To expand on this, $\sigma$ is a quantity that has a bearing on the change in the potential barrier in time $\Delta t^\prime$. A higher value of $\sigma$ implies that this change is smaller and vice-versa.
As the value of $\sigma$ is decreased from $100$, we first observe a gradual increase in maximum power (due to greater work output). However, below $\sigma = 4$, we see a sharp decline in the value of maximum power. It should be noted that the lower the value of $\sigma$, the greater the change in the height of the potential barrier in $\Delta t^\prime$. Consequently a larger value of $r$ is required to extract useful (positive) work, as we have seen in Fig.~\ref{fig:Work-speed}. As we decrease $\sigma$ beyond the maximum value of maximum power, the $r$ needed to achieve useful work increases greatly. Thus a much larger time $\Delta t$ is need to get thermalized and this causes the sudden decrease of maximum power.

The efficiency at maximum power for finite time quantum heat engines is usually given by Curzon-Ahlborn efficiency~\cite{Curzon1975, Abe2011, Abah2012, Wang2012, Dann2020, Johal2021b}. However, in our work we cannot include the Curzon-Ahlborn value in our discussion as we have considered, for simplicity, instantaneous thermalization in steps 2 and 4 (labeled in Fig.~\ref{fig:CSC}) of the quantum Stirling cycle.



\begin{figure}
\begin{center}
\includegraphics[scale=0.5]{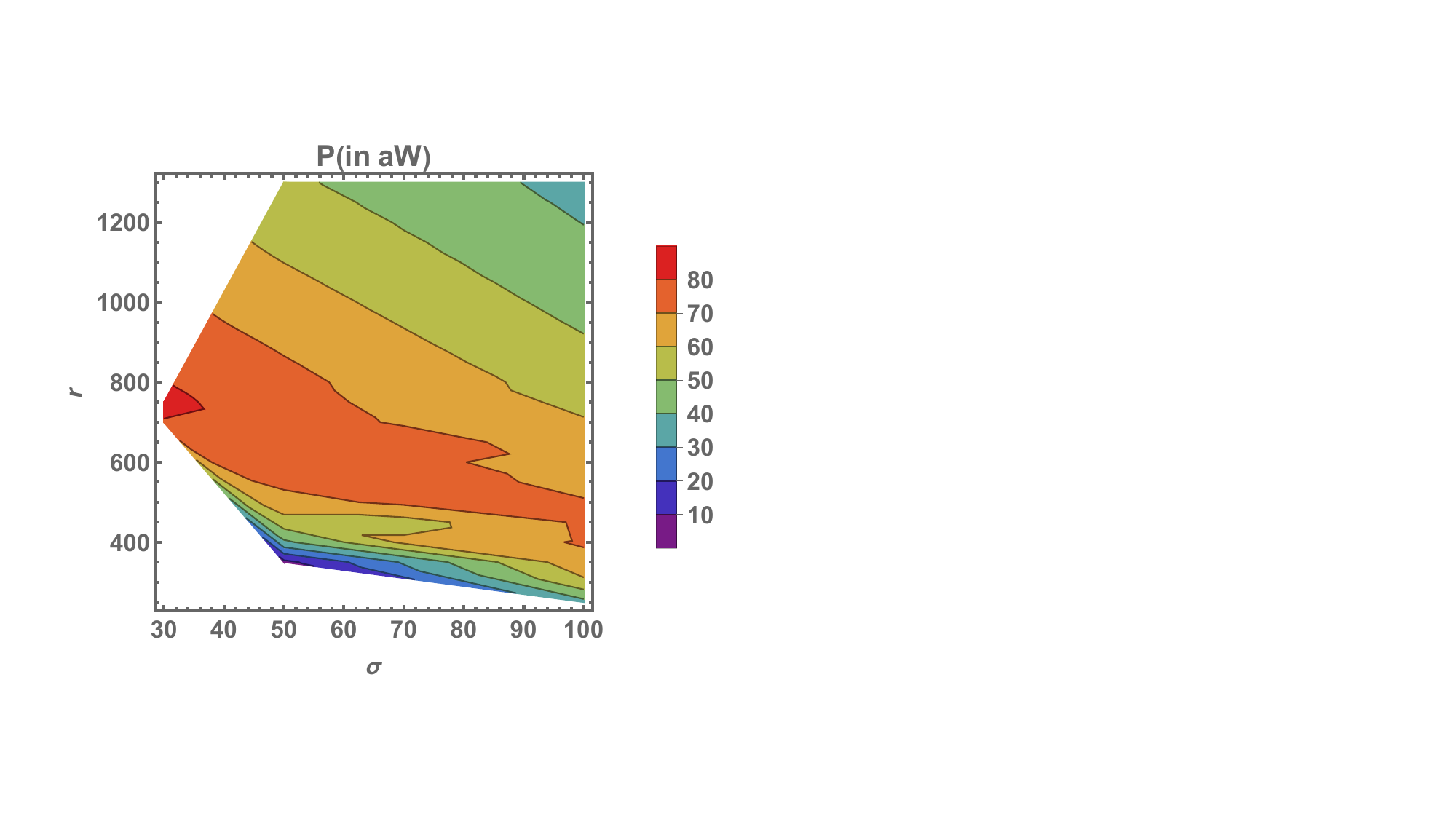}
\end{center}
\caption{Plot of output power as a function of $r$ and $\sigma$ in the regime of slow insertion and removal of the potential barrier. For a constant value of $\sigma$, larger values of output power are obtained for intermediate values of $r$. In the upper half of the plot, for a constant $r$, a decrease in $\sigma$ results in an increase in output power. In the lower half of the plot, however, the opposite is observed.}
\label{fig:contour}
\end{figure}

The summary of our results can perhaps be best expressed using a contour plot of output power as a function of $r$ and $\sigma$ (or equivalently $\Delta \alpha$) as shown in Fig.~\ref{fig:contour}. The dependence of output power on $r$ and $\sigma$ is demonstrated in the regime of relatively slow insertion and removal of the potential barrier. In this regime, for a constant value of $\sigma$, the power increases with an increase in $r$ and then decreases. The larger values of power are achieved at intermediate values of $r$. This behavior mirrors what we have previously seen in Fig.\ref{fig:p-speed}. If, on the other hand, $r$ is kept constant at a high value (roughly in the top half of the plot) and $\sigma$ is decreased progressively, we see an increase of the power. This aspect is related to the tail end of the plot (on the right side of the peak) in Fig.\ref{fig:max-p-speed}. The idea that can be understood from this plot is that if $r$ is kept constant at a relatively low value (roughly the lower half of the plot) and $\sigma$ is decreased, we see a decrease of the power output.

In the regime of very fast insertion and removal of the potential barrier or for $\sigma< 4$,  we should see a different behavior. Here power decreases as we decrease $\sigma$ and this is related to the behavior seen in Fig.\ref{fig:max-p-speed} on the left side of the peak. However, positive output power in this region requires very large values of $r$ $(\gg 10^6)$ and many points are difficult to obtain for the purpose of plotting. Hence we have concentrated on the regime of slow driving speed. In particular, certain points in the white regions the plot represent values of $r$ and $\sigma$ where there is no output power.

Let us recapitulate our results. The quantum Stirling cycle is realized in two ways. In the first approach, the length of the time step is varied and the change in the height of the potential barrier is kept constant. In this way a longer time step corresponds to a longer interaction time with the thermal bath and a slower cycle. We see that a quantum Stirling engine does not yield useful work if operated very fast. In the quasistatic regime, the efficiency approaches the Carnot efficiency. However, the output power is higher when the quantum Stirling cycle is operated faster than the quasistatic rate of operation, although the efficiency progressively declines. In the second approach, the time-step is constant but the change in the height of the potential barrier is varied. A bigger change corresponds to a faster cycle. We see that in this case the maximum output power initially increases as the cycle is operated faster, but after a certain speed it decreases steeply.

\section{Discussions and Conclusion}
\label{sec:conc}
We have modeled a finite-time Stirling engine at low temperatures, based on a particle in a box. In our model, in each time step an instantaneous adiabatic change in the potential barrier occurs, followed by a period of thermalization. We have discussed the dependence of work, efficiency, and power on both the duration of the thermalization step and the adiabatic change of the potential barrier. In the slow operating regime, the work extracted approaches the limiting value of the work obtained in a quasistatic quantum Stirling cycle. Also, the efficiency of the engine approaches the Carnot value in this regime. For very fast operation of the cycle the system does not function as a heat engine. The output power for insertion and removal steps is found to attain its maximum value not in the quasistatic regime but at an intermediate operating speed. For a constant duration of thermalization, there is an optimal change of potential barrier for which the maximum output power is highest. This investigation is important to obtain the practical implementations of this quantum heat engine or similar quantum heat engines. 

The model of interaction with the thermal bath that has been used here can be used to analyze other quantum heat engines in finite time that operate by inserting and removing a potential barrier. The quantum Szilard engine, for example, has been explored in the context of a harmonic potential well~\cite{Davies2021} instead of the traditional potential box. However, an implementation in practice requires an understanding of the operation in finite time, involving the interaction with the thermal bath which our method can provide. Other future directions of this work can involve potentials that might provide additional benefits and asymmetric insertion and removal of the barrier potential. For example, a single-conduction-band effective-mass envelope function for confined electrons in two-dimensional GaAs quantum double dots can conceivably be used to mimic the double-well potential of a quantum Stirling engine~\cite{Hu2000}. In GaAs, the application of appropriate electric voltages over nanoscale electrodes, lithographically engineered, produces a suitable confining potential by creating a depletion area in a two-dimensional electron-gas. The confinement potential and the barrier height are defined by controllable parameters and could be suitable for realizing this quantum heat engine. Double quantum wells have also been realized using InAs/ GaSb quantum dots~\cite{Pal2015, Mueller2015, Karalic2016, Karalic2017}. The double-potential-well structure appropriate for a quantum Stirling engine can also be created using a symmetric two electron Si quantum double dot by manipulating the mixing of conduction band valleys~\cite{Tariq2022}.

The  mass has been chosen as $m_e$ and the size of the box has been chosen as 20 nm for all the simulations as they correspond to the mass and length scales suitable for practical applications. The role played by the size of the potential box was previously studied and discussed in Ref.~\cite{Thomas2019}. A large-size box corresponds to the classical limit. In this limit, the energy levels approach a continuous energy spectrum. At length scales smaller than the Compton wavelength, relativistic effects enter the picture and the analysis would require a different kind of treatment.
\section*{Acknowledgments}
We thank Jukka P. Pekola for very useful discussions.  This work was supported by the John Templeton Foundation, Grant No. 61835. G.T. acknowledges support from the Academy of Finland, QTF Centre of Excellence (Grant No. 312057).

%

\end{document}